\documentclass[seceq,preprint]{ptptex}

\notypesetlogo                       
\preprintnumber[3cm]{arXiv:0711.1611v3}

\title{
Comments on Chemical Potentials in AdS/CFT%
}

\author{
Shin \textsc{Nakamura}\footnote{E-mail: nakamura@hanyang.ac.kr}%
}

\inst{
Department of Physics, BK21 Program Division,\\
Hanyang University, Seoul 133-791, Korea\\
and\\
Center for Quantum Spacetime, Sogang University,\\
Seoul 121-742, Korea
}

\abst{
We propose a method for identifying holographic chemical potentials of conserved charges. The guiding principle is the consistency of the identification with the thermodynamic relations and the Legendre transformation. We consider the baryon-charge chemical potential as an example, and explain why the degree of freedom of the constant shift of the bulk $U(1)$ gauge field is absent when the Legendre transformation is well-defined. The method proposed here suggests that the definition of the chemical potential may be more complicated compared with the case of localized charge if we have a nontrivial charge distribution along the radial direction of the bulk geometry. 
}

\begin{document}

\maketitle

\section{Introduction}
The application of gauge/gravity correspondence to quark-hadron physics has recently attracted much attention. In particular, AdS/CFT at finite baryon density is important since there is a technical difficulty called the ``sign problem'' (see for example, Ref. \citen{Lattice}) in lattice QCD when introducing the finite baryon chemical potential.
Holographic descriptions of systems at finite baryon density have been studied in a number of papers and many interesting results have been obtained.\cite{KSZ,HT,ParSah,NSSY,KMMMT,Bergman,DGKS,UBC,KSZ-2,NSSY-2,Par,Kyusyu,KB,MMMT,EKR,Mats}$^{,}$\footnote{Other related references include Ref. \citen{baryon}.}
However, there are issues that are still under debate. One of them is related to holographic definition of the chemical potential. 

It is known that the global flavor symmetry is promoted to the local (gauge) symmetry on the flavor brane in the gravity-dual side.
The $U(1)_{\mbox{\scriptsize B}}$ symmetry, which is the diagonal part of the global flavor symmetry, corresponds to the $U(1)$ gauge symmetry on the flavor brane.
Thus, we can naturally identify the ``electric charge'' on the flavor brane as a bulk counterpart of the $U(1)_{\mbox{\scriptsize B}}$ charge.
However, we have several ways of identifying the baryon chemical potential with the bulk field. It is natural to relate the non-normalizable mode of $A_{0}$, the zeroth component of the $U(1)$ gauge field on the flavor brane, to the chemical potential since it is the conjugate field to the electric charge. However, we need to establish the dictionary in a gauge-invariant way.\footnote{
We consider finite-temperature systems in this paper where the Euclidean time direction is compactified using the periodicity of the inverse temperature $\beta=1/T$. The gauge transformation we are considering is one that respects the periodicity.}
There are at least two methods of defining a gauge-invariant quantity related to the boundary value of $A_{0}$:
\begin{enumerate}
  \item $\mu=\frac{1}{\beta}\int^{\beta}_{0}dt A_{0}|_{\rho=\infty}$
\ \ \ (Definition 1),
  \item $\mu=\int^{\infty}_{\rho_{\mbox{\scriptsize min}}}d\rho F_{\rho 0}$
\ \ \ \ \ \ \ (Definition 2),
\end{enumerate}
where $F_{\rho 0}\equiv \partial_{\rho}A_{0}-\partial_{0}A_{\rho}$ and $\rho$ is the radial coordinate of the bulk geometry whose boundary is located at $\rho=\infty$. $\rho_{\mbox{\scriptsize min}}$ is the point where the flavor brane terminates inside the bulk. Depending on the setup and the dynamics, the brane may terminate at the horizon of the bulk geometry (black hole embeddings) or elsewhere (Minkowski embeddings). The first definition is also gauge invariant in finite-temperature systems since the Euclidean time direction is compactified.
If we assume a static configuration of the $U(1)$ gauge field, the above quantities are reduced to $A_{0}(\infty)$ and $A_{0}(\infty)-A_{0}(\rho_{\mbox{\scriptsize min}})$, respectively. The latter can also be equivalent to $A_{0}(\infty)$ if we choose $A_{0}(\rho_{\mbox{\scriptsize min}})=0$; however, a crucial difference between them is whether or not we allow the constant shift of $A_{0}$ as a physically meaningful degree of freedom.

Definition 2, or Definition 1 with $A_{0}(\rho_{\mbox{\scriptsize min}})=0$, has been used in Refs.~\citen{HT,ParSah,NSSY,KMMMT,KSZ-2} and \citen{NSSY-2} while Definition 1 has recently been employed successfully in Refs. \citen{Bergman,DGKS,UBC,Kyusyu,KB,MMMT} and \citen{Mats} (and also in the pioneering Ref. \citen{KSZ}). 
The difference between the two definitions disappears on black hole embeddings because we must set $A_{0}=0$ at the horizon where the Euclidean time circle shrinks to zero, while the difference can survive in principle on Minkowski embeddings.
However, the degree of freedom of the constant shift of $A_{0}$ has also been fixed at finite baryon-charge density even on Minkowski embeddings (see for example, Ref. \citen{Bergman}). Thus, a natural question arises: when and why is the degree of freedom of the constant shift forbidden? 

In this paper, we will answer this question in a model-independent way. The key point is the consistency of the definition of the chemical potential with the Legendre transformation and the thermodynamic relations. In \S \ref{Legendre}, we demonstrate how the identification of the chemical potential is related to the thermodynamic potentials and the Legendre transformation by using a toy model to visualize the problem. In \S \ref{proposal}, we reinterpret this demonstration.
We will see that Definition 2 (and its generalized version) is naturally selected, at least at finite charge density where the Legendre transformation is well-defined. We find that the absence of the constant-shift degree of freedom of $A_{0}$ results from the fact that the grand potential in the gravity dual contains two terms: one of them is the source term, which corresponds to (the expectation value of) the charge density in YM theory, and the other is the charge projection operator, which is explained in \S \ref{proposal}. Therefore, the absence of the constant-shift degree of freedom of $A_{0}$ holds in general when the model has these two terms. In \S \ref{proposal}, we also propose a general method for defining the baryon (and other) chemical potentials in general setups such as those containing the mass of baryons. In the discussion section, we consider a case where we have a nontrivial charge distribution along the $\rho$ direction. We point out that the definition of the chemical potential may be more complicated in the presence of a nontrivial charge distribution in the bulk. 

Unfortunately, the argument presented in this paper does not apply to the case where the free energy is independent of the chemical potential (if such a sector exists). Indeed, all the known sectors where Definition 1 plays an important role are charge-less Minkowski embeddings on which this property is realized. \cite{Bergman,DGKS,UBC,Kyusyu,KB,MMMT,Mats} In this sense, we are not going to dispute the validity of Definition 1 in this special sector in the present work. 


\section{Consistency of Legendre transformation}
\label{Legendre}

We consider a system where the $U(1)$ charges are present on the flavor branes. Here, we ignore the dynamics of the charges and we assume they are massless and localized at $\rho=\rho_{\mbox{\scriptsize min}}$. The model may not be very close to phenomenologically realistic setups, however it is sufficiently close to allow us to observe an important feature related to the chemical potential. 

The total Lagrangian\footnote{
We employ the probe approximation where the back reaction to the bulk geometry from the flavor brane is ignored, and the bulk Lagrangian is omitted since it does not affect the discussion in this paper under the approximation. The Lagrangian should be understood as being renormalized, although we do not write the counterterms explicitly.}
of the system is given by
\begin{eqnarray}
\int^{\infty}_{\rho_{\mbox{\scriptsize min}}}d\rho{\cal L}
=\int^{\infty}_{\rho_{\mbox{\scriptsize min}}}d\rho {\cal L}_{\mbox{\scriptsize DBI}}-QA_{0}(\rho_{\mbox{\scriptsize min}}),
\end{eqnarray}
where ${\cal L}_{\mbox{\scriptsize DBI}}$ is the DBI Lagrangian of the flavor branes. We have assumed translational (and rotational) symmetry of the system; all the bulk fields depend only on $\rho$, and integrals over the other directions have already been evaluated.
The amount of $U(1)_{\mbox{\scriptsize B}}$ charge is $Q$, which is understood to be thermal expectation value if we are in the grand canonical ensemble, whereas it is a control parameter in the canonical ensemble. The minus sign in front of the source term originates from the fact that the charge induced on the flavor brane is always opposite to the quark charge inserted in the D3-branes (where the YM theory is applicable) in the picture before replacing the D3-branes with the near-horizon geometry.

We need to define the on-shell Lagrangian to specify the thermodynamic potentials. Here the meaning of on-shell is that the total Lagrangian satisfies the equations of motion, including 
\begin{eqnarray}
\partial_{\rho}\frac{\partial{\cal L}}{\partial A'_{0}}
=\frac{\partial{\cal L}}{\partial A_{0}}=-Q\delta(\rho-\rho_{\mbox{\scriptsize min}}),
\label{eom}
\end{eqnarray}
as well as the boundary conditions in such a way that all the charges are on the brane:
\begin{eqnarray}
\left. \frac{\partial{\cal L}}{\partial A'_{0}}
\right|_{\infty}=-Q,
\ \ \ \
\left. \frac{\partial{\cal L}}{\partial A'_{0}}
\right|_{\rho_{\mbox{\scriptsize min}}}=0.
\label{bound-cond}
\end{eqnarray}
We ignore the scalar fields on the brane since they do not contribute within the context of the present section (see Appendix \ref{scalar}).

\subsection{Chemical potential as $A_{0}(\infty)$}
\label{wrong-mu}

Let us start with Definition 1 where we regard $A_{0}(\infty)$ as the chemical potential. The grand potential $\Omega$, which is consistent with the thermodynamic relation $Q=-\partial \Omega/\partial \mu$ in this case, is given (up to $\mu$-independent terms) by\footnote{
We omit $|_{\mbox{\scriptsize on-shell}}$ from the next equation.}
\begin{eqnarray}
\Omega=\left. \int d\rho {\cal L}\right|_{\mbox{\scriptsize on-shell}}.
\label{omega-1}
\end{eqnarray}
Let us verify its consistency explicitly:
\begin{eqnarray}
\delta\Omega
&=&\int d\rho \left\{\partial_{\rho}\left[\frac{\partial{\cal L}}{\partial A'_{0}}\delta A_{0}\right]
-\left[\partial_{\rho}\frac{\partial{\cal L}}{\partial A'_{0}}
 -\frac{\partial{\cal L}}{\partial A_{0}} \right]\delta A_{0}
\right\}.
\nonumber \\
&=&
-Q\delta A_{0}(\infty),
\label{delta-omega-1}
\end{eqnarray}
where we have used the equations of motion and the boundary conditions (\ref{bound-cond}). We have derived the thermodynamic relation
$Q=-\partial \Omega/\partial \mu$
without imposing any further constraint on $A_{0}(\rho_{\mbox{\scriptsize min}})$.

Let us perform a Legendre transformation on $\Omega$ to define the Helmholtz free energy $F$:
\begin{eqnarray}
F=\Omega+\mu Q
=\int d\rho {\cal L}_{\mbox{\scriptsize DBI}}-QA_{0}(\rho_{\mbox{\scriptsize min}}) 
+Q A_{0}(\infty).
\label{F-0}
\end{eqnarray}
If the above construction is consistent, we need to derive the correct thermodynamic relation $\partial F/\partial Q=\mu$. Let us examine explicitly whether or not this is the case: 
\begin{eqnarray}
\frac{\partial F}{\partial Q}
&=&A_{0}(\infty)-A_{0}(\rho_{\mbox{\scriptsize min}})
\nonumber \\
&&+\int d\rho \frac{\partial{\cal L}}{\partial A'_{0}}
\frac{\partial A'_{0}}{\partial Q}
+Q\frac{\partial}{\partial Q}
\left\{A_{0}(\infty)-A_{0}(\rho_{\mbox{\scriptsize min}})
\right\}.
\end{eqnarray}
Here, the second line simplifies to zero since:
\begin{eqnarray}
\int d\rho \frac{\partial{\cal L}}{\partial A'_{0}}
\frac{\partial A'_{0}}{\partial Q}
&=&\int d\rho 
\left\{
\partial_{\rho}
\left[
\frac{\partial {\cal L}_{\mbox{\scriptsize DBI}}}{\partial A'_{0}} \frac{\partial A_{0}}{\partial Q}
\right] 
-
\left[ 
\partial_{\rho}
\frac{\partial {\cal L}_{\mbox{\scriptsize DBI}}}{\partial A'_{0}}
\right]
\frac{\partial A_{0}}{\partial Q}
\right\}
\nonumber \\
&=&
-Q\frac{\partial A_{0}(\infty)}{\partial Q}
+Q\frac{\partial A_{0}(\rho_{\mbox{\scriptsize min}})}{\partial Q},
\end{eqnarray}
where we have used the equations of motion, boundary conditions (\ref{bound-cond}) and the fact that $\frac{\partial {\cal L}_{\mbox{\scriptsize DBI}}}{\partial A'_{0}}=\frac{\partial {\cal L}}{\partial A'_{0}}$.
Therefore, we obtain
\begin{eqnarray}
\frac{\partial F}{\partial Q}
=A_{0}(\infty)- A_{0}(\rho_{\mbox{\scriptsize min}}),
\label{delFdelQ}
\end{eqnarray}
where the second term is absent at the starting point. If we follow the above procedure, the thermodynamic relations and the Legendre transformation do not close under the chemical potential given by Definition 1.

\subsection{Chemical potential as $A_{0}(\infty)- A_{0}(\rho_{\mbox{\scriptsize min}})$}
\label{correct-mu}

Now, let us start with Definition 2 of the chemical potential:
\begin{eqnarray}
\mu=A_{0}(\infty)- A_{0}(\rho_{\mbox{\scriptsize min}}).
\label{true-chem-0}
\end{eqnarray}
A grand potential that is consistent under this definition is
\begin{eqnarray}
\Omega=\int d\rho {\cal L}_{\mbox{\scriptsize DBI}}.
\label{omega-2}
\end{eqnarray}
Notice that we have removed the source term from the new $\Omega$. Let us verify the consistency:
\begin{eqnarray}
\delta\Omega
&=&\int d\rho \left\{\partial_{\rho}\left[\frac{\partial{\cal L}_{\mbox{\scriptsize DBI}}}{\partial A'_{0}}\delta A_{0}\right]
-
\left[
\partial_{\rho}\frac{\partial{\cal L}_{\mbox{\scriptsize DBI}}}{\partial A'_{0}}
\right]
\delta A_{0}
\right\}.
\nonumber \\
&=&
-Q\left\{
\delta A_{0}(\infty)-\delta A_{0}(\rho_{\mbox{\scriptsize min}})
\right\},
\label{delta-omega-2}
\end{eqnarray}
where we have used the same on-shell conditions (\ref{eom}) and (\ref{bound-cond}). Equation (\ref{delta-omega-2}) gives the correct thermodynamic relation $\partial \Omega/\partial \mu=-Q$ under the present definition.

Let us perform a Legendre transformation on the above $\Omega$ to obtain the Helmholtz free energy:
\begin{eqnarray}
F=
\int d\rho {\cal L}_{\mbox{\scriptsize DBI}}
+Q\left\{A_{0}(\infty)-A_{0}(\rho_{\mbox{\scriptsize min}})\right\}.
\label{FF-0}
\end{eqnarray}
Interestingly, the free energy (\ref{FF-0}) is exactly the same as Eq. (\ref{F-0}). We have already seen that Eq. (\ref{F-0}) has a consistent thermodynamic relation (\ref{delFdelQ}) under the dictionary (\ref{true-chem-0}).

\section{A method for defining the chemical potential}
\label{proposal}

We have seen in the previous section that the consistency with the thermodynamic relations and the Legendre transformation may indicate how to uniquely identify the chemical potential. Let us reorganize the results of the previous section to clarify matters.

We have obtained the same Helmholtz free energy starting with the different definitions of the chemical potential, one of which was selected on the basis of the consistency with the thermodynamic relation $\partial F/\partial Q=\mu$. This means that the Helmholtz free energy plays a fundamental role in the  definition of the chemical potential in our formalism. Indeed, the canonical ensemble is a better starting point for us than the grand canonical ensemble, since the correspondence between the $U(1)_{\mbox{\scriptsize B}}$ charge and the $U(1)$ charge on the flavor brane is clearer than that between the chemical potential and $A_{0}$. These observations suggest that we should start with the free energy (\ref{F-0}) or (\ref{FF-0}): 
\begin{eqnarray}
F=\int d\rho {\cal L}+QA_{0}(\infty).
\label{FF-1}
\end{eqnarray}
The first term is the total Lagrangian of the system. 
The second term is simply the {\em charge projection operator} originally introduced into black hole thermodynamics to define the Helmholtz free energy \cite{ChargeProjection,CP2}.

Let us remind ourselves of what the charge projection operator is. If we start with the total Lagrangian, its variation is given by
\begin{eqnarray}
\delta L=(\mbox{\rm term giving the equations of motion}) 
-Q\delta A_{0}(\infty),
\end{eqnarray}
where the last contribution originates from the boundary term. However, we need to control the charge $Q$ rather than $A_{0}$ since we are in the canonical ensemble. If we add the charge projection operator to the total Lagrangian, the variation becomes
\begin{eqnarray}
\delta (L+QA_{0})=(\mbox{\rm term giving the equations of motion}) 
+(\delta Q) A_{0}(\infty);
\end{eqnarray}
thus, we can employ the same equations of motion while holding the charge fixed. The point is that we need to choose an appropriate expression of the free energy depending on how we control the parameter.

We can reinterpret the results of the previous section in this context. For example, Eq. (\ref{delta-omega-1}) shows that we obtain the equations of motion by extremizing\footnote{
The on-shell constraint is removed from Eqs. (\ref{omega-1}) and (\ref{omega-2}) when we discuss the extremization.
}  
the grand potential (\ref{omega-1}) with $A_{0}(\infty)$ kept fixed but without fixing $A_{0}(\rho_{\mbox{\scriptsize min}})$. Alternatively, we obtain the same equations of motion by extremizing another grand potential (\ref{omega-2}) by fixing both $A_{0}(\rho_{\mbox{\scriptsize min}})$ and $A_{0}(\infty)$. We have chosen the appropriate grand potential depending on how we control the boundary conditions. We have (at least) two possible choices at this stage. However, we have found that only one of them, given in \S \ref{correct-mu}, is consistently connected to the unique expression of the Helmholtz free energy (\ref{FF-1}) by the Legendre transformation.

The method for defining the chemical potential is now clear:
\begin{enumerate}
  \item Find the charge projection operator with respect to the conserved charge under consideration.
  \item Add the charge projection operator to the total (on-shell) Lagrangian of the system to define the Helmholtz free energy.
  \item Differentiate the Helmholtz free energy with respect to the charge to find the conjugate chemical potential.
  \item Perform the Legendre transformation, if necessary, to switch to the grand canonical ensemble.  
\end{enumerate}

Let us examine how this works in more general setups. We consider, as an example, the Sakai-Sugimoto model with massive charged sources, which is studied in Ref. \citen{Bergman}. The total Lagrangian added to the charge projection operator is simply the Helmholtz free energy employed in Ref. \citen{Bergman}:
\begin{eqnarray}
F=\int d\rho {\cal L}_{\mbox{\scriptsize DBI}}+ L_{\mbox{\scriptsize source}}+QA_{0}(\infty),
\end{eqnarray}
where $L_{\mbox{\scriptsize source}}$ is the Lagrangian of the baryon-charged objects, which consists of their mass contribution ($L_{\mbox{\scriptsize mass}}$) and the source ($-QA_{0}(\rho_{\mbox{\scriptsize min}})$). The chemical potential obtained by differentiating the free energy with respect to the charge is
\begin{eqnarray}
\mu
=A_{0}(\infty)-A_{0}(\rho_{\mbox{\scriptsize min}})
+
\frac{\partial L_{\mbox{\scriptsize mass}}}{\partial Q},
\end{eqnarray}
after taking account of the force-balance condition \cite{Bergman}. The last term is the mass of the baryon-charged object, which is now naturally incorporated into the definition of the baryon chemical potential. The above definition, which is a variant of Definition 2, does {\em not} contain the degree of freedom of the constant shift of $A_{0}$.

Indeed, we can show that the degree of freedom of the constant shift is always absent when $L_{\mbox{\scriptsize source}}$ contains the charged source term balanced with the charge projection operator. This explains the absence of the constant-shift degree of freedom from the model-independent definition of the chemical potential.

However, there is a caveat. 
The method proposed above does not work if $\partial F/\partial Q$ is singular.\footnote{
We are not referring to the singularity at phase transition points. A phase transition is defined as a jump between different branches of the solutions of the equations of motion. 
Our concern is whether or not $\partial F/\partial Q$ is well-defined within a single branch of the solutions.
} 
For example, a sector where the amount of charge remains zero regardless of the chemical potential has been considered in Refs. \citen{Bergman,DGKS,UBC,Kyusyu,KB,MMMT} and \citen{Mats}. This is a Minkowski embedding without the charge, and we call it the ``trivial sector'' in this paper. Obviously, $\partial F/\partial Q$ is singular in such a sector and our method does not apply. Therefore, we do not claim that our results apply to the trivial sector in the present work; all the statements in this paper apply only to the case where $\partial F/\partial Q$ is well-defined.

\section{Discussion}
\label{discussions}

We have seen that the natural definition of the chemical potential is Definition 2 (or its generalization) rather than Definition 1 when $\partial F/\partial Q$ and the Legendre transformation are well-defined. 
A crucial point is that the degree of freedom of the constant shift of $A_{0}$ does not exist except for the very special case where $\partial F/\partial Q$ is singular.
We have also proposed a general method for defining the chemical potential in terms of the bulk quantities.

We now add a few comments on systems with nontrivial charge distribution along the $\rho$ direction.\footnote{
Such a case has been studied in Ref. \citen{UBC}.
} 
If the charge is not localized at a particular value of $\rho$, the definition of the chemical potential becomes more complicated. For example, the toy model we have considered in \S \ref{Legendre} can be generalized in the following way. Suppose that the total Lagrangian is given by
\begin{eqnarray}
\int^{\infty}_{\rho_{\mbox{\scriptsize min}}}d\rho{\cal L}
=\int^{\infty}_{\rho_{\mbox{\scriptsize min}}}d\rho 
\left\{
{\cal L}_{\mbox{\scriptsize DBI}}-q(\rho)A_{0}(\rho)
\right\},
\end{eqnarray}
where $q(\rho)$ is the charge density along the $\rho$ direction, which satisfies $\int d\rho\: q(\rho)=Q$. The charge projection operator we need to add is still $QA_{0}(\infty)$ since the boundary term still results in the same total charge inside the system by virtue of the Gauss law. Then the Helmholtz free energy is given by the on-shell value of
\begin{eqnarray}
F
=\int^{\infty}_{\rho_{\mbox{\scriptsize min}}}d\rho 
\left\{
{\cal L}_{\mbox{\scriptsize DBI}}-q(\rho)A_{0}(\rho)
\right\}+QA_{0}(\infty),
\end{eqnarray}
and the chemical potential is given by
\begin{eqnarray}
\mu
&=&A_{0}(\infty)
-\int^{\infty}_{\rho_{\mbox{\scriptsize min}}}d\rho 
\frac{\partial q(\rho)}{\partial Q} A_{0}(\rho)
\nonumber \\
&=&
\int^{\infty}_{\rho_{\mbox{\scriptsize min}}}dr 
\frac{\partial q(r)}{\partial Q}
\int^{\infty}_{r}d\rho F_{\rho 0},
\label{general-chem}
\end{eqnarray}
which is again written in terms of the field strength.\footnote{
Equation (\ref{general-chem}) can be formally interpreted as the hypothetical work against the electric field necessary, to bring the unit charge from the boundary to accomplish the new charge distribution on top of the old one. 
} The response of the distribution to the variation of the total charge, $\partial q(r)/\partial Q$, must be determined by the dynamics. It is certainly worthwhile investigating how this identification works in various general setups.

This discussion is rather general and it applies to any chemical potential in principle. Thus, it is also interesting to consider the isospin chemical potential \cite{Par,EKR,isospin} in holographic setups using the method outlined in this paper. Since mesons can carry the isospin charge, we can discuss them within the framework of the (nonabelian) DBI theory of flavor branes without introducing any extra objects such as baryon vertices or fundamental strings; the finite isospin system may be a suitable test ground\footnote{
Of course, we need to consider both baryon and isospin chemical potentials in phenomenologically realistic setups.} for the proposed method.

\section*{Acknowledgements}
The author would like to thank Sang-Jin Sin, Tetsuo Hatsuda, Yunseok Seo, Youngman Kim and Sangmin Lee for discussions and comments. The author thanks the hospitality of the Elementary Particle Theory Group at Kyushu University where part of the present work was carried out.
This work was supported by KOSEF Grant
R01-2004-000-10520-0 and the SRC Program of the KOSEF through the
Center for Quantum Spacetime of Sogang University (grant number
R11-2005-021).

\appendix
\section{Scalar-Field Dependence} 
\label{scalar}

In the main text, we have ignored the scalar fields on the flavor brane which may contribute to the variation of the thermodynamic potentials when we vary $Q$ or $\mu$. We show that the contribution indeed vanishes \cite{Bergman}. Suppose that the DBI Lagrangian contains a scalar field $y$. Then, the additional contribution to the variation of the free energies that may originate from the $y$ field is 
\begin{eqnarray}
\int^{\infty}_{\rho_{\mbox{\scriptsize min}}}d\rho
\frac{\partial {\cal L}_{\mbox{\scriptsize DBI}}(y')}{\partial y'}\frac{\partial y'}{\partial \mu}
=
\left.
{\rm (const)}\frac{\partial y}{\partial \mu}
\right|^{\infty}_{\rho_{\mbox{\scriptsize min}}},
\label{y-contri}
\end{eqnarray}
where we have used the equation of motion for $y$:
$\partial {\cal L}_{\mbox{\scriptsize DBI}}(y')/\partial y'={\rm const.}$
Here, $\partial y/\partial \mu |_{\infty}$ is zero because the boundary value of $y$ determines another parameter of the theory such as the current quark mass, which is kept fixed under the variation of the chemical potential. Then, (\ref{y-contri}) indicates the variation oroginates from only the $y(\rho_{\mbox{\scriptsize min}})$ dependence of the action. However, this is zero because of the force-balance condition of the flavor brane along the $y$ direction at $\rho=\rho_{\mbox{\scriptsize min}}$. The same logic applies to differentiation with respect to the charge. 

%

\end{document}